# Six Potential Game-Changers in Cyber Security: Towards Priorities in Cyber Science and Engineering

**Alexander Kott**
US Army Research Laboratory
2800 Powder Mill Road
Adelphi, Maryland    20783
United States

alexander.kott1.civ@mail.mil

**Ananthram Swami**
US Army Research Laboratory
2800 Powder Mill Road
Adelphi, Maryland    20783
United States

ananthram.swami.civ@mail.mil

**Patrick McDaniel**
Computer Science and Engineering
Pennsylvania State University
360A IST Building
University Park, PA    16802
United States

mcdaniel@cse.psu.edu

## 1.0    INTRODUCTION

The fields of study encompassed by cyber science and engineering are broad and poorly defined at this time. As national governments and research communities increase their recognition of the importance, urgency and technical richness of these disciplines, a question of priorities arises: what specific sub-areas of research should be the foci of attention and funding?  In this paper we point to an approach to answering this question. We present results of a recent workshop that explored possible "game-changers" or disruptive changers that might occur in cyber security within the next 15 years. We suggest that such game-changers may be useful in focusing attention of research communities on high-priority topics. Indeed, if a drastic, important change is likely to occur, should we not focus our research efforts on the nature and ramifications of the phenomena pertaining to that change?

In the last few years, terms like "Science of Cyber," or "Science of Cyber Security," or "Cyber Science" have been appearing in use with growing frequency. For example, the US Department of Defense defined "Cyber Science" as a high priority for its science and technology investments [1], and the National Security Agency has been exploring the nature of the "science of cybersecurity" in its publications [2]. There are a number of reasons for the rise of interest in science of cyber. A key motivator is the recognition that development of cyber technologies is handicapped by the lack of scientific understanding of the cyber phenomena, particularly the fundamental laws, theories, and theoretically-grounded and empirically validated models [3]. The US President's National Science and Technology Council [4] stressed the lack of such fundamental knowledge, and its importance.

While there are many in the cyber security community who agree with the need for science of cyber—whether it merits an exalted title of a new science or should be seen merely as a distinct field of research within one or more of established sciences—the exact nature of the new science, its scope and boundaries remain rather unclear.

As proposed in [5] [6], the primary objects of research in cyber security are:

- Attacker(s) $A$ along with the attackers' networks and tools (especially malware) and techniques $T_a$
- Defender $D$ along with the defender's defensive tools and techniques $T_d$, and operational assets, networks and systems $N_d$
- Policy $P$, a set of defender's assertions or requirements about what event should and should not happen. To simplify, we may focus on cyber incidents $I$: events that should not happen.

This allows one to summarize the classification of major problem groups in cyber security. Below, for each subclass, example of common problems in cyber security research and practice are added, for the sake of illustration:

- $T_d, T_a, I \rightarrow N_d$, e.g., synthesis of network's structure, planning and anticipation of network's behavior, assessing and anticipating network's security properties
- $N_d, T_a, I \rightarrow T_d$, e.g., design of defensive tools, algorithms, planning and control of defender's course of action, assessing and anticipating the efficacy of defense
- $T_d, N_d, I \rightarrow T_a$, e.g., identification of structure of attacker's code or infrastructure, discovery, anticipation and wargaming of attacker' actions, anticipating the efficacy of attacker's actions
- $T_d, N_d, T_a \rightarrow I$, e.g., detection of intrusions that have occurred, anticipation of intrusions that will occur

Clearly, within the broad outlines of such problem classes, one finds endless opportunities for specific research topics. The question of priorities among these topics, particularly from the perspective of military R&D funding managers, deserves plenty of pondering.

The thoughts offered in this article, although primarily the responsibility of the authors, have been triggered and shaped to a large extent by a long-term cyber security research planning meeting convened by the Cyber Collaborative Research Alliance (Cyber CRA – see cra.psu.edu, also: http://www.arl.army.mil/www/default.cfm?page=1417). Sponsored by the US Army Research Laboratory, Cyber CRA is a collaborative effort of the US Government researchers and a consortium of universities, led by the Pennsylvania State University, to advance the theoretical foundations of cyber science.

The overall objective of the Cyber Security CRA is to develop a fundamental understanding of cyber phenomena, including aspects of human attackers, cyber defenders, and end users, so that fundamental laws, theories, and theoretically grounded and empirically validated models can be applied to a broad range of Army domains, applications, and environments. This multi-year, basic research program aims to develop and advance the state of the art of Cyber Security in the following areas:

- The Risk Research Area seeks to develop theories and models that relate fundamental properties and features of dynamic risk assessment algorithms to the fundamental properties of dynamic cyber threats, Army's networks, and defensive mechanisms. Intuitively, this discipline seeks to understand what assets are at risk and what is the impact of those risks being realized.

- The Detection Research Area seeks to develop theories and models that relate properties and capabilities of cyber threat detection and recognition processes/mechanisms to properties of a malicious activity, and of properties of Army networks. Detection falls into a larger scientific exploration of the identification and understanding of "situational awareness".

- The Agility Research Area seeks to develop theories and models to support planning and control of cyber maneuver (i.e., "maneuver" in the space of network characteristics and topologies) that would describe how control and end-state of the maneuver are influenced by fundamental properties of threats, such as might be rapidly inferred from limited observations of a new, recently observed threat. This discipline seeks to understand what remediation actions are possible and effective for a given adversary or attack, as well as estimating their costs.

Each of the above Research Areas must take into take into account that the human is central to cyber security, as a significant part of the problem as well as the solution. In the cyber security domain there are three human elements: the user/Soldier, attacker/adversary and the analyst/defender, acting as individuals or as group(s).

To inform its research program planning, the Cyber CRA decided to explore the likely trends and changes in cyber security technologies over the next fifteen years. The two-day exploratory meeting took place in Scottsdale, AZ, in May 2014. The meeting emphasized eliciting the industry-driven perspective, as opposed to the more commonly explored views of academia.

A total of 26 organizations, the majority of them from the industry, were represented at the meeting. In addition to corporate organizations—ranging from major multinational developers of hardware and software to smaller cyber-security firms—several academics with strong industry ties, and several US Department of Defense laboratories joined the meeting.

## 2.0 THE SIX POTENTIAL GAME-CHANGERS

In our view, the meeting yielded six major themes that may be seen as reasonably likely game-changers in the cyber technology landscape in the next ten to twenty years. We discuss them below, in the following order. The first two are potential changes in the cyber environment; the next two are technology trends that are already in effect but might grow to a qualitatively greater strength; and the last two are the less likely but potentially highly influential breakthroughs in cyber technology capabilities.

### 2.1 New Computing Paradigms

The first game changer is the potential emergence of new computing paradigms, ranging from nano-computing, quantum computing, DNA-based computing, etc., to new system and network architectures, such as cloud architectures taken to extremes. These novel paradigms might in the span of just a few years drastically change the market and technology of cyber security, making most current technologies obsolete.

Nanocomputing and analog computing, for example in sensor applications, may require qualitatively new types of exploits and defenses. Quantum computing and networking are already feeding lively disputes about their alleged inherent security versus alleged opportunities for hacking. Wearable computing sharpens the challenges of usability and user perceptions and behaviors, and privacy concerns. These challenges will be even greater when a computing device and network augments and even invades the human, as in the case of prosthetics, exoskeletons, and brain-computer interfaces. Internet of Things (IoT) greatly increases opportunities for cyber attacks but might also offer new approaches to defense, e.g., collective defense of multiple devices.

Wearable and IoT computing will likely create demand for custom devices, designed and fabricated to specific customer's specifications. A device may even be designed and fabricated by the user himself—compare this to the current surge of popularity of 3D printing and small hobby computers. Each unique device will require a customized cyber defense mechanism.

Biologically inspired approaches to cyber security, such as artificial immune systems, may attract growing interest, especially as they offer promises of autonomous self-healing and adaptation to previously unknown threats. However, they may also bring inherent unpredictability and complexity of their behaviors.

An example of an innovative, biologically-inspired computational and communication paradigm is the Gaian Database (Fig. 1), a dynamic distributed federated database [7]. The US Army Research Laboratory (ARL) and the UK Ministry of Defense (MoD) partly funded the research and development of the Gaian DB, and ARL researchers contributed to the experimental development of the technology [8]. In 2013, it was deployed and successfully demonstrated at the NATO Intelligence Fusion Centre (NIFC) at RAF Molesworth, UK. Gaian is particularly useful for connecting and integrating highly heterogeneous, mobile computing devices, ranging from a large-scale data center to thousands of handsets—reminiscent of a tactical Army environment.

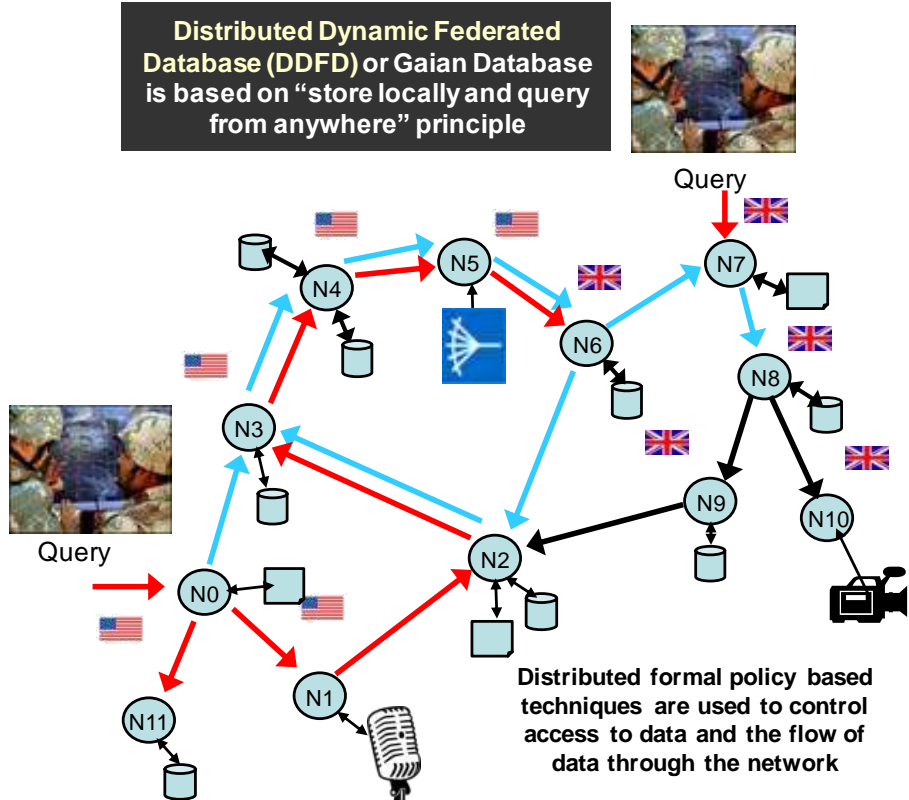

**Figure 1: Gaian platform is particularly useful for connecting and integrating highly heterogeneous, mobile computing devices in coalition environments.**

Self organizing bio-inspired Gaian offers a number of features that pose new challenges to both cyber defenders and cyber attackers. It is based on the collaboration model with no central point of control; hides data and database heterogeneity and compensates for data & schema mismatch; provides data location transparency but include provenance; effectively copes with dynamic modification of network and data topology [7]. Distributed formal policy based techniques are used to control access to data and the flow of data through the network—very useful in coalition warfare where coalition partners may have different policies on access and use of the data they collect and manage.

The example hints at some of the challenges of dealing with novel computing paradigms. It is necessary, in parallel, with developing a new computing and communication paradigm, to perform a systematic analysis of new devices and architectures from the cyber security perspective. This must be done proactively, before the paradigm is productized, implemented, and deployed, at which time it may become a victim of innovative cyber attacker with unpredictable ramifications to the users. However, scientific foundations, engineering approaches, standards and methodologies that could support such a concurrent analysis are currently lacking.

## 2.2 Reaching the Threshold of Complexity

The second game-changer is the potential crossing of a threshold of network complexity that brings us into a new territory beyond the limits of conventional system manageability and human comprehension. This qualitative phase change in technology complexity—enormous size, connectivity, interdependence, heterogeneity and dynamics—might defeat conventional approaches of current science and engineering disciplines.

This trend could be engendered in part by the near-inevitable Internet of Things, proliferation of cyber-physical systems, and enormous growth of social networks throughout currently under-involved countries. Such a discontinuity might invalidate much of the assumption underpinning current cyber security. For example, a participant of our meeting expressed this sentiment, "Our products are already far too complex, so complex that nobody in our corporation can possibly fully understand them, and this just keeps getting worse." Arguably, increased complexity of our systems enhances the adversary's opportunities for asymmetric attacks.

Of course, predictions of collapse of internet due to its excessive size and complexity have been made in the past, and have been proven spectacularly misguided. The ever-growing complexity of Internet and other highly complex systems is exceeded only by their remarkable adaptability and resiliency. However, it can be observed that the size and diversity of the computing systems is often an advantage for those who would defend it. Here, however, we are concerned not about exceeding the limits of Internet, but rather the limits of human comprehension (and cyber defense) of such complex systems.

A new level of development in the sciences of complexity, and corresponding qualitatively novel advances in tools of complex system analysis, modeling and synthesis, will be required to cope with the change. Such sciences and engineering tools would need to address enormous multitudes of explicit and emergent interactions and highly nonlinear and dynamic dependencies within the system. These theories and tools would also deal with cyber strengths, vulnerabilities and control aspects of the system that encompasses a wide diversity of individual nodes, including human nodes, with their relatively unpredictable physical and cognitive behaviors. New approaches will be needed to measurements of complexity, to finding or predicting tipping points. New hierarchical constructs that reduce unstable and unanticipated interactions may emerge that would help deal with complexity.

Cyber security is particularly challenged by the complex, multi-genre nature of networks that combine several distinct genres—networks of physical resources, communication networks, information networks, and social and cognitive networks. Study of systems such as multi-genre networks is relatively new; instead, it is customary in research and engineering literature to focus on a view of a network comprised of homogeneous elements, (e.g., a network of communication devices, or a network of social beings). Yet, most if not all real-world networks are multi-genre—it is hard to find any real system of a significant complexity that does not include a combination of interconnected physical elements, communication devices and channels, data collections, and human users forming an integrated, inter-dependent whole. The ARL Network Science CTA (www.ns-cta.org) focus is on modeling, influencing and understanding the co-evolution of such interacting mutli-genre networks [9].

When we consider multiple genres of networks, the total number of links that connect any two elements grows significantly. Perrow [10] explains that catastrophic failures of systems emerge from high complexity of links which lead to interactions that the system's designer cannot anticipate and guard against. System's safety precautions can be defeated by hidden paths, incomprehensible to the designer because the links are so numerous, heterogeneous, and often implicit. Greater connectivity we recognize in a multi-genre network helps us see more of the overall network's complexity, and hence the potential influences on its robustness and resiliency [11].

However, numerous challenges arise in dealing with impacts of network complexity on cyber security. For example, the cyber research community lacks insights into stability, controllability and observability of networks, especially those of very large scale and of multi-genre nature. Little is understood about the impact of social-cognitive links and information-element links on the overall network complexity [11a]. From an engineering perspective, it would be highly desirable to have design approaches that would enable dynamic, on-demand segmentation of a potentially unstable network into subnets with limited and strictly controlled connectivity between them. Yet, such designs and design tools are currently absent.

## 2.3 Big Data Analytics

The third game-changer is the emergence of big data analytics—predictive and highly autonomous—in cyber security. This is a current trend, still immature but already noticeable and influential. Cyber analytics does have a potential to become truly big (world-scale), predictive or anticipative in its analytical conclusions (i.e., it would anticipate new cyber threats and malicious activities in actionable details and time frames), and highly autonomous, i.e., able to operate with little or no participation of human cyber analysts. If the potential of this trend is realized, cyber defense will acquire a new degree of potency.

Because some of the most promising applications of big data analytics are found in medical and biomedical research [12], analogies between cyber security and such field as epidemiology are useful. For example, Gil, et al [13], describe an approach inspired by genetic epidemiology where features of a computer host—such as the network services active on a computer—are statistically linked to the kinds of threats to which that host is likely to be susceptible. This is similar to the tools used in genetics to identify statistical associations between mutations and diseases, and like in the case of genetic epidemiology relies on availability of large volumes of data, especially when rare conditions are of interest.

Much of the power of big data analytics is likely to derive from aggregating and correlating a broad range of highly heterogeneous data. As the history of the field of information fusion has demonstrated, the related challenges are very formidable. When heterogeneity of the data is further complicated by their noise, incompleteness and massive scale characteristic of cyber data—the challenges become even greater [14]. Much work is also needed in developing new analytical algorithms for inferring well-hidden and possibly deception-protected information from massive collections of heterogeneous, rapidly evolving data. And while cyber defenders need anticipative results from such analytic algorithms, it is not clear to which extent such anticipation is even theoretically feasible.

Threats, however, are also likely to evolve, possibly utilizing a form of such analytics for their own purposes. The adversary may use his own big data analytics to uncover information about our vulnerabilities and defenses. In addition, the adversary may use our own big data analytics, in at least two ways. He might be able to access parts of our data storages or analytic facilities, gaining a remarkably rich view of our capabilities and their limits. He may also provide our data collecting devices with deceptive, misleading, or probing data. Means to anticipate such actions will be needed and their impact estimated [14a, 14b].

Other challenges in this area include the lack of methods and standards for efficiently preserving and tagging the collected, highly heterogeneous information with meta-data useful for future processing for a variety of purposes. Some early work on fully homomorphic encryption in the US-UK-ITA holds promise for computationally feasible processing of data while preserving privacy [15] [16]. Also lacking are standards, languages and recommended practices for bringing and aggregating information from diverse organizations, especially in a multi-national environments, where different nations may have different constraints on release and use of information. Another challenge in this area is the now-ubiquitous Big Data problem: collection of data from different layers of the OS and protocol stack, and across the network leads to an explosion of data; processing of this data is made difficult by the inability to communicate to a fusion center or to even fit into memory for local processing [17]; mining the low-dimension information-rich manifold remains a challenge

[18].

## 2.4   Resilient Self-Adaptation

The fourth game-changer would be potential breakthroughs in approaches based on resilient self-adaptation, where cyber security derives largely from the system's agility, moving target defense (e.g., Fig. 2), cyber maneuvering, autonomous response, self-healing and other such autonomous or semi-autonomous behaviors [19]. Because it is unlikely that practically invulnerable, impenetrable systems and networks can be built consistently and economically, it might be beneficial to shift a significant fraction of design resources from reducing vulnerabilities to increasing resiliency. Here, a resilient system is a system that may experience a major loss of capability due to a cyber attack, but recovers these capabilities sufficiently rapidly and fully so that the overall mission of the system proceeds successfully. Note the focus on a mission; resiliency is meaningful only in the context of a mission.

For example, promising results exist for approaches where a software residing on a host like a mobile phone performs self-healing—by applying patches or other forms of self-rewriting its code—in response to its own abnormal behavior it detects [20]. Recovery at the network scale can benefit from approaches such as reported by Mell and Harang [21] that perform a post-intrusion triage to highlight the most likely infected nodes, so that the remediation efforts would produce the highest degree of resiliency.

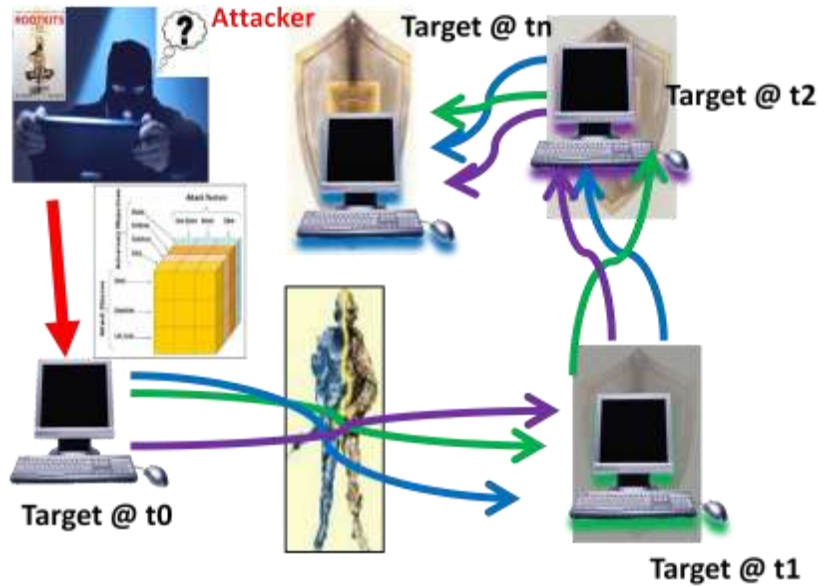

**Figure 2: Approaches called Moving Target Defense or Cyber Maneuver aim to confuse the attacker regarding the true structure and functions of the defended network.)**

However, in spite of the fact that a few of these approaches are already implemented by researchers and by hardware and software providers, overall this space of approaches remains to be researched and exploited. Their broad applicability, feasibility and ultimate value remain to be proven and at this time are far from certain. If successful, however, this trend could drastically change the current cyber security industry.

Novel design and analysis methods, along with new architectural patterns would need to emerge to make design for resiliency an accepted and successful practice. Elements of suitable approaches could be gleaned from work on autonomous designs [21a, 21b] and adaptive, dynamic planning [21c]. Still, challenges of such designs are manifold. For example, effective autonomous self-adaptation or self-healing seems to call for a degree of machine intelligence that may be far ahead of what can be currently imagined. It may also increase the system's complexity, which in itself is a concern and a potential multiplier of vulnerabilities as we discuss in this article. Given the diversity and unpredictability of complex attacks and their circumstances, practical resiliency can only be probabilistic in nature—not a comforting thought for the operators of the system. Besides, a resilient self-adaptation of the system is likely to conflict with the operators' desire for understandable and predictable behaviors and outcomes. Further, self-adaptation brings with it challenges of resistance to external control, which may not be desirable.

From the perspective of practical engineering, it would help to stabilize the use of terminology that currently ranges somewhat interchangeably from Moving Target Defense to Cyber Maneuver [22] to Agility to Rapid Recovery, etc. These are probably best assigned to specific phases of defense, e.g., prior to compromise (perhaps MTD), during the compromise (e.g., Cyber Maneuver), etc.

### 2.5 Design of Mixed Trusted Systems

The fifth game-changer is the potential emergence of new design methods for security-minded, flexible, modifiable systems that combine and accommodate un-trusted and legacy hardware and software, as well as clean-slate components. The sources of untrusted software and hardware (whether new or legacy) are numerous, ranging from dubious supply chains to ever-growing complexity of new components to convergence of business systems with personal systems and across the continually expanding multi-national environments. This ubiquity of diverse computing infrastructure and the necessary reliance on less advanced and secure networks require new means to enhance cyber security without necessarily fixing the vulnerabilities in the conventional sense.

Related ideas are explored, for example [23], where a trusted core element is introduced within an autonomic ad hoc sensor networks when some nodes are untrusted, corrupted or malicious. Trust-based schemes, implemented by a trusted core, result in significant improvement in the state estimation procedure, and the interplay between estimation and trust updates quickly isolates malicious nodes. Alternatively, a trust management protocol, as opposed to a trust core, could be used to apply trust-based intrusion detection to assess the trustworthiness and maliciousness of sensor nodes [24] [25].

To be broadly applicable, this game-changer would entail creation of major, qualitatively new design methodologies and tools that enable synthesis of complex systems by, for example, reinforcing untrusted components with clear-slate and highly trusted "braces" (compare to seismic resistant insulating bearings, etc.). Such designs would also include components that could be rapidly and inexpensively modified to defend against new threats as they are discovered. Formal methods would have to see a breakthrough, or yet unknown semi-formal but highly reliable methods would need to emerge.

The new design tools would need to unobtrusively encourage developers toward more secure and verifiable systems and architectures. Such tools would not merely ensure that the design is secure, but will also tell the designers and maintainers of the system why the design is secure (e.g., stating explicitly security assumptions, rationale, proof, arguments, constraints, etc.), and where the design can be modified without compromising its security. The resulting designs must balance security against usability, meeting the needs of the mission and cognitive demands of users. The tools would need to account for emergent properties of accidental or unexpected composition and reconfiguration of components. To support such approaches, the precision of security-related standards and requirements would need to improve significantly.

## 2.6 Active and Proactive Responses

The sixth game-changer involves the potential emergence of approaches to active and proactive responses to threat sources, strategy-oriented approaches, offence-based techniques, alternative security postures, deception and psychological mechanisms. Currently, little is understood regarding the shape or possibility of such methods, especially in view of legal and policy uncertainties surrounding the cyber security in general, and any active responses to cyber threats in particular. If such developments were to accelerate, they could change the calculus of attacker-defender relations, and engender an entirely new generation of technologies. The long history and extensive knowledge of strategy and tactics in conventional conflicts might offer important insights, but also could be misleading and limiting in some cases.

To enable and guide the developments in such approaches, many assumptions, often implicit and long ingrained in our understanding of dealing with adversaries—from crime fighting and law enforcement to armed conflict—may need to be explicitly examined and perhaps questioned. Issues are broad and inadequately understood in the context of cyber security. Examples include: moral standards versus national and international legal standards; non-territorial nature of the cyber space; impact of cultural differences and biases; interactions of cyber and physical spaces and activities; public versus private spaces; definitions of strengths and weaknesses our ours and our adversaries.

Like in conventional conflicts, such approaches might aim to defeat the adversary's ability and will to wage the conflict by holding the adversary at risk. The risk to the adversary may take multiple forms: economic costs of cyber attacks; possibility of retaliation (cyber or conventional); concerns of falling for a deception; internal dissent and infighting, etc.

Human factors are of critical importance in cyber security as they are in all forms of warfare. For example, situational awareness of the adversary determines the adversary's ability to conduct effective attacks against the friendly networks while defending his resources against our counter-actions. Although [26] there are significant differences between situational awareness in conventional and cyber conflicts, similarities are sufficiently strong to enable transfer of lessons and techniques learned in conventional conflicts to cyber security.

In all cases, advanced capabilities of the adversary have to be assumed. Regardless of details, such approaches will benefit from greater situational awareness, e.g., knowing our own and adversary's architectures and infrastructure, and sensing capabilities. Also needed would be languages that clearly and precisely articulate the situation; cultural intelligence and adversary modeling; and deep insights into individual and collective cognitive processes of the adversary.

## 3.0 CONCLUSIONS AND RECOMMENDATIONS

Considering the relatively long time horizon, it would be foolhardy to suggest that all these six game-changers, or even any one of them, will emerge as envisioned here, if at all. It is reasonable, however, to suggest that paying special attention to the related developments—or lack thereof—will benefit cyber security researchers, technologists, and their funding organizations. Furthermore, to prioritize efforts towards cyber science and engineering, it would be appropriate to give extra weight to those topics that contribute to potential game changers, such as those we outline in this paper.

Specific recommendations for advancement of cyber science and engineering can be offered for each of the six game-changers. With regard to the potential emergence of new computing paradigms systematic, proactive analysis of security of newly emerging devices and architectures should be given consideration, along with analytical approaches, methodologies and standards generalized for a broad range of paradigms. Also needed is synthesis of customized cyber defense mechanism, perhaps such as self-evolving artificial immune systems, along with means to predict their security properties.

With respect to reaching a threshold of network complexity research is needed on controllability, observability, and stability of large, multi-genre (i.e., including social-cognitive and informational aspects) networks; also important are means for near-real time measurements of complexity properties and for finding tipping points; approaches to controlling non-determinism; as well as engineering design paradigms and tools that can dynamically semi-isolate network segments and control the behaviors of and between the segments.

To leverage the ongoing trend towards the big data analytics, the research community should seek means for fusion of heterogeneous, noisy and incomplete information on massive scales; and new analytical algorithms for inferring well-hidden and possibly deception-protected information from massive collections of rapidly evolving data. In addition, methods and standards are necessary for tagging the data with meta-information, and standards, languages and recommended practices for bringing data from diverse organizations, often with divergent policies on access and utilization of the data.

Approaches based on resilient self-adaptation will benefit from a clear taxonomy of methods and concepts of moving target defense, cyber maneuver and such, and from leveraging applicable, mature research disciplines, e.g., cryptography and key distribution for MTD, control theory for cyber maneuver and recovery.  Priority should be given to novel methods for analysis and design centered on resiliency properties of systems.

Potential emergence of new design methods and tools that synthesize complex systems from trusted and untrusted components will require theoretical foundations, currently absent, for system theory with components at heterogeneous levels of security and trust. To overcome currently slow progress in formal methods, consideration should be given to possible use of semi-formal and approximate methods.

Finally, with respect to active and proactive responses, approaches that employ forms of deception or concealment—that is the means to mislead the adversary regarding our networks and his effects on our networks—are likely to be less controversial and easier to implement than others in this broad and challenging space of techniques.

## ACKNOWLEDGEMENTS

The authors are grateful to Andrew Toth of the US Army Research Laboratory and Dr. Cliff Wang of the Army Research Office for information about several of the research examples mentioned in this paper and to John MacLeod for editing and correcting the manuscript.